\documentclass[12pt]{article}

\usepackage{amsmath}
\usepackage{amssymb}
\usepackage{amsfonts}
\usepackage{lscape}

\begin{document}

\fontsize{12}{6mm}\selectfont
\setlength{\baselineskip}{2em}

$~$\\[.35in]
\newcommand{\dss}{\displaystyle}
\newcommand{\raro}{\rightarrow}
\newcommand{\be}{\begin{equation}}

\def\sech{\mbox{\rm sech}}
\def\sn{\mbox{\rm sn}}
\def\dn{\mbox{\rm dn}}
\thispagestyle{empty}

\begin{center}
{\Large\bf A Chiral Schwinger Model,}  \\    [2mm]
{\Large\bf its Constraint Structure and}  \\    [2mm]
{\Large\bf Applications to its Quantization}   \\   [2mm]
\end{center}

\vspace{1cm}
\begin{center}
{\bf Paul Bracken}                        \\
{\bf Department of Mathematics,} \\
{\bf University of Texas,} \\
{\bf Edinburg, TX  }  \\
{78541-2999}
\end{center}

\vspace{3cm}
\begin{abstract}
The Jackiw-Rajaraman version of the chiral
Schwinger model is studied as a function of the
renormalization parameter. The constraints are
obtained and they are used to carry out canonical
quantization of the model by means of Dirac 
brackets. By introducing an additional scalar field,
it is shown that the model can be made gauge
invariant. The gauge invariant model is
quantized by establishing a pair of gauge
fixing constraints in order that the method of
Dirac can be used.
\end{abstract}

\vspace{2mm}

\vspace{2mm}

\newpage
\begin{center}
{\bf 1. Introduction.}
\end{center}

Quantization of a theory, its effects on the 
classical symmetries and the
mechanisms for mass generation of particles
in a quantum field theory are
subjects that continue to be of interest.
Many important effects are already visible
in the context of a smaller model with
several degrees of freedom. Electrodynamics
and its nonabelian extensions in one-space
and one-time dimension with massless fermions
is of great interest for many reasons, one
of which is that the quantization of the theory
can be studied from various points of view.
The Schwinger model describes a massless
Dirac field in two-dimensions with both
chiral components coupled to a $U(1)$ gauge 
field {\bf [1]}. The Jackiw-Rajaraman model {\bf [2,3]} 
or chiral Schwinger model is related to this 
model but has coupling to only one chiral component,
and it is found to depend on a regularization
parameter. Thus there is an anomaly in the
singly-coupled model, which cancels against
a similar but sign reversed anomaly in the
doubly-coupled Schwinger model.
Moreover, this parameter, although
arbitrary, has a very significant effect
on the structure of the model, especially the
constraint equations  and the nature 
of its field equations. Quantum field theories with 
gauge couplings to chiral fermions have the
property that there is an anomalous nonconservation
of the gauge current. This is an interesting 
characteristic in itself, but of even more significance
is that gauge invariance may be lost as well
as the existence of a consistent theory {\bf [4]}.
Since most of the
interest in gauge theories in general arises
from the fact that they are both
renormalizable and unitary, this is a serious
problem. Consequently,
to ensure that renormalizability and
unitarity are not threatened, the 
structure of a theory may be modified
or extended. This is evident in the case of
gauge theories in which the gauge group
is adjusted so that the fermion content of
the theory satisfies a specified rule,
for example, the number of quarks equals the
number of leptons in the model. 
Consequently, there can be an 
anomaly for gauge symmetry in a subsector
of the theory, but all the individual 
contributions must cancel.

The purpose here is to investigate the chiral
Schwinger model and show how it 
corresponds to the bosonized chiral
Schwinger model in one-space and one-time 
dimension which is due to Jackiw and
Rajaraman {\bf [2]}. It will be seen that the 
structure of the model, especially the
structure of the constraints, depends
on the value assigned to the regularization
parameter. The structure of the system of
constraints will be obtained for several
different cases of this parameter and discussed
in detail. It is shown how the theory
corresponding to the classical Hamiltonian
can be quantized  based on the Dirac bracket {\bf [5]}.
The study of the constraints is important
as far as path integral quantization is
concerned, especially when gauge fixing
constraints must be added to convert a
set of first class constraints into a set
of second class constraints {\bf [6,7]}. The topic of
constraints has been of interest recently {\bf [8,9]}

Gauge invariance
may be lost in this process, but it will be shown 
in a case in which this takes place that,
by including a type of Wess-Zumino term {\bf [10]},
gauge invariance can be restored.
To accomplish this however a new field
must be introduced. Although this effectively
enlarges the Hilbert space, gauge invariance
is retained and the field appears in the Hamiltonian
in a way analogous to the scalar boson field 
already present. This is in contrast to an
alternative procedure which is to use
BRST quantization of a gauge invariant theory {\bf [11]}. 
To do this, the theory is rewritten as a
quantum system that possesses a generalized gauge 
invariance, and require that the Hilbert space
of the gauge invariant theory be enlarged.
In this formulation, the gauge-invariant theory
replaces the gauge transformation by a BRST
transformation. This transformation mixes
operators having different statistics, and
as with the Wess-Zumino field, the corresponding
Hilbert space is enlarged.

It will be seen that gauge invariance is restored
using this Wess-Zumino term when the parameter is one.
Two gauge fixing constraints are introduced
into the theory which serve to establish a gauge.
Using all the constraints, a quantization
of the theory can be performed {\bf [12,13]}. Finally a path integral
quantization will be outlined at the end {\bf [14]}.

\begin{center}
{\bf 2. Introduction and Properties of the Model}
\end{center}

The Lagrangian density for the chiral Schwinger model is
given explicitly by
$$
{\cal L}_S  = - \frac{1}{4} F^{\mu \nu} F_{\mu \nu}
+ \bar{\psi} [ i \not{\partial} + e
\sqrt{\pi} \not{A} (1 + i \gamma_5 )] \psi,
\eqno(2.1)
$$
where $\gamma_5 = i \gamma^0 \gamma^1$.
At the classical level, the Lagrangian (2.1)
is invariant under the local gauge transformations
$$
A_{\mu}(x) \rightarrow A_{\mu} (x) + \partial_{\mu} \alpha (x),
\qquad
\psi(x) \rightarrow e^{2 ie \sqrt{\pi} \alpha(x) P_{+}} \psi (x),
\qquad
\bar{\psi} (x) \rightarrow \bar{\psi} (x)
e^{-2 i e \sqrt{\pi} \alpha (x) P_{-}},
$$
where $P_{\pm} = \frac{1}{2} ( 1 + i \gamma_5)$.
The fermion determinant for this two-dimensional
system can be evaluated in closed form and yields
an effective action of the form
$$
S_e = \int dt \, dx \, \{ - \frac{1}{4} F_{\mu \nu}
F^{\mu \nu} + \frac{e^2}{2} A_{\mu}
( a g^{\mu \nu} - ( g^{\mu \alpha} - \epsilon^{\mu \alpha})
\frac{\partial_{\alpha} \partial_{\beta}}{\Box}
(g^{\beta \nu} - \epsilon^{\beta \nu})) A_{\nu} \}.
\eqno(2.2)
$$
The quantity $a$ in (2.2) is a constant which is
not uniquely determined by the different
procedures for calculating the fermionic
determinant. Its value would be fixed by gauge
invariance were it not for the fact that the
model has an anomaly {\bf [15]}. However, $a$ may be
allowed to be arbitrary, but the domain of $a$ will be 
restricted to a particular subset of values.
The two cases $a>1$ and $a=1$ will
be of interest here and studied separately,
and we will take $\hbar =1$ in what follows.

It will now be shown that an auxiliary scalar field
$\varphi (x)$ can be introduced into the formalism, 
which links (2.1) to the bosonized version of the model,
by introducing a path integral with respect
to the scalar field which can be done in closed 
form as follows
$$
\exp [ i S_e (A) ] = \int {\cal D} \varphi \exp
[ i S ( A, \varphi ) ].
\eqno(2.3)
$$
The action on the right-hand side of (2.3) is
modified from (2.2) to include the new scalar
field $\varphi$  as
$$
S (A, \varphi) = \int \, dt \, dx \, {\cal L} ( A, \varphi ),
\eqno(2.4)
$$
where the Lagrangian with the scalar field is
$$
{\cal L} ( A, \varphi)
= - \frac{1}{4} F_{\mu \nu} F^{\mu \nu} +
\frac{1}{2} ( \partial_{\mu} \varphi)(\partial^{\mu}
\varphi) + e ( g^{\mu \nu} - \epsilon^{\mu \nu} )
A_{\nu} \partial_{\mu} \varphi + \frac{1}{2}
a e^2 A_{\mu} A^{\mu}.
\eqno(2.5)
$$
In (2.4), $g^{\mu \nu}$ is the Lorentz metric,
$$
g^{\mu \nu} = \left(
\begin{array}{cc}
1  &  0  \\
0  & -1  \\
\end{array}   \right)
= g_{\mu \nu},
$$
and $\epsilon^{\mu \nu} =- \epsilon^{\nu \mu}$.

To see that (2.4) is the bosonized version of the
fermion action, note that the $\varphi$
integration is independent of the field $A_{\mu}$.
By using (2.4) and (2.5) in (2.3), then 
separating the $\varphi$-dependent terms in
the path integral as
$$
\int \, D \varphi \,
\exp [ i \int d^2 x \, [ - \frac{1}{4} F_{\mu \nu}
F^{\mu \nu} + \frac{1}{2} ( \partial_{\mu} \varphi)(
\partial^{\mu} \varphi) + e (g^{\mu \nu} - \epsilon^{\mu \nu})
A_{\nu} \partial_{\mu} \varphi + \frac{1}{2} a e^2
A_{\mu} A^{\mu}]]
$$
$$
= \exp [ i \int d^2 x \, \{ - \frac{1}{4} F_{\mu \nu}
F^{\mu \nu} + \frac{1}{2} a e^2 A_{\mu}A^{\mu} \} ]
\int D \varphi \, \exp [i \int d^2 x \,
\{ - \frac{1}{2} \varphi \Box \varphi - e
( g^{\mu \nu} - \epsilon^{\mu \nu} ) \partial_{\mu} A_{\nu} 
\varphi \}].
$$
Now the integral with respect to $\varphi$ can be 
done by completing the square and up
to an irrelevant multiplicative determinant factor, this
matches $S_e$ in (2.2). The presence of the scalar
field in (2.5) also allows another interpretation
of the parameter $a$, namely, it reflects the
degree of bosonization in the model.
Once the action has been determined, as in (2.4),
it is straightforward to determine the field
equations for both the boson field $\varphi$ and 
the vector potential $A_{\mu}$ using the
Euler-Lagrange equations {\bf [16]}
$$
\partial_{\mu} \frac{\partial  L}{\partial_{\mu} Q_r}
- \frac{\partial L}{\partial Q_r} = 0,
$$
where $Q_r$ stands for either of the two fields
$\varphi$ or $A_{\mu}$. The following two equations 
are obtained from (2.5) {\bf [17]}
$$
\Box \varphi + e ( g^{\mu \nu} - \epsilon^{\mu \nu} ) A_{\nu}
=0,
\eqno(2.6)
$$
$$
\partial_{\mu} F^{\mu \nu} +e ( g^{\nu \alpha}
- \epsilon^{\nu \alpha} ) \partial_{\alpha} \varphi
+ a e^2 A^{\nu} = 0.
\eqno(2.7)
$$
Note that $\varphi$ can be obtained explicitly in
terms of $A_{\nu}$ from (2.6)
$$
\varphi = - e (g^{\mu \nu} - \epsilon^{\mu \nu})
\frac{\partial_{\mu} A_{\nu}}{\Box}.
\eqno(2.8)
$$
Substituting $\varphi$ into (2.7), there results 
the expression
$$
\partial_{\mu} F^{\mu \nu} + a e^2 A^{\nu}
- e^2 ( g^{\nu \alpha} - \epsilon^{\nu \alpha})
\frac{\partial_{\alpha} \partial_{\beta}}{\Box}
(g^{\beta \mu} - \epsilon^{\beta \mu} ) A_{\mu} =0.
\eqno(2.9)
$$
It will be shown that a solution to system (2.6)-(2.7)
is determined by taking
$$
A^{\mu} = - \frac{1}{ae} [ \partial^{\mu} \varphi
+ (1 - a) \epsilon^{\mu \nu} \partial_{\nu} \varphi
- a \epsilon^{\mu \nu} \partial_{\nu} h ],
\eqno(2.10)
$$
where $h$ is an arbitrary function that satisfies
the wave equation
$$
\Box h = 0,
\eqno(2.11)
$$
and the function $\varphi + h$ satsifies the 
Klein-Gordon equation
$$
\Box ( \varphi + h) + \frac{e^2 a^2}{a-1} ( \varphi + h) =0.
\eqno(2.12)
$$
To show that (2.6) is satisfied, we calculate 
$g^{\mu \nu} \partial_{\mu} A_{\nu}$  using the antisymmetry
of $\epsilon^{\mu \nu}$
$$
g^{\mu \nu} \partial_{\mu} A_{\nu}
= - \frac{1}{ae} [ \Box \varphi + (1- a)
\epsilon^{\mu \tau} \partial_{\mu}
\partial_{\tau} \varphi - a \epsilon^{\mu \tau}
\partial_{\mu} \partial_{\tau} h ]
= - \frac{1}{ae} \Box \varphi,
$$
and moreover, we obtain
$$
\epsilon^{\mu \nu} \partial_{\mu} A_{\nu}
= - \frac{1}{ae} [ (1- a) \epsilon^{\mu \nu} \epsilon_{\nu}^{\tau}
\partial_{\mu} \partial_{\tau} \varphi
- a \epsilon^{\mu \nu} \epsilon_{\nu}^{\tau}
\partial_{\mu} \partial_{\tau} h ]
$$
$$
= - \frac{1}{ae} [ ( 1- a) g^{s \tau } \epsilon^{\mu \nu}
\epsilon_{\nu s} \partial_{\mu} \partial_{\tau} \varphi
- a g^{\sigma s} \epsilon^{\mu \nu} \epsilon_{\nu s}
\partial_{\mu} \partial_{\tau} h ]
$$
$$
= - \frac{1}{ae} [ ( 1 -a) g^{\tau \mu} \partial_{\mu}
\partial_{\tau} \varphi - a g^{\tau \mu}
\partial_{\mu} \partial_{\tau} h]
$$
$$
=- \frac{1}{ae} [ (1-a) \Box \varphi - a \Box h ].
$$
Therefore,
$$
( g^{\mu \nu} - \epsilon^{\mu \nu} ) \partial_{\mu} 
A_{\nu} = - \frac{1}{ae} \Box \varphi + \frac{1}{ae}
(1-a) \Box \varphi = - \frac{1}{e} \Box \varphi.
$$
This is exactly (2.6). Similarly, we calculate
$$
\Box A^{\mu} = - \frac{1}{ae} [ \partial^{\mu} 
\varphi + ( 1 -a) \epsilon^{\mu \nu} \partial_{\nu}
\Box \varphi ],
$$
and 
$$
\partial^{\mu} \partial_{\nu} A^{\nu} = -
\frac{1}{ae} \Box \partial^{\mu} \varphi.
$$
Using these, we find that
$$
\Box A^{\mu} - \partial^{\mu} \partial_{\nu} A^{\nu}
+ a e^2 A^{\mu}
$$
$$
= - \frac{(1-a)}{ae} \epsilon^{\mu \nu} \partial_{\nu}
\Box \varphi - e \partial^{\mu} \varphi
- e ( 1-a) \epsilon^{\mu \nu} \partial_{\nu} \varphi
+ a e \,  \epsilon^{\mu \nu} \partial_{\nu} h
$$
$$
= \frac{(1-a)}{ae} \epsilon^{\mu \nu} \partial_{\nu}
( \Box ( \varphi + h) + \frac{a e^2}{a-1} (\varphi +h))
- e ( g^{\mu \nu} + \epsilon^{\mu \nu}) \partial_{\nu}
\varphi.
$$
If it is required that $\sigma = \varphi + h$ 
satisfy the additional equation
$$
\Box \sigma + \frac{a^2 e^2}{a-1} \sigma = 0,
\eqno(2.13)
$$
then (2.7) holds. When $a \neq 1$, equation (2.13) is a Klein-Gordon
equation which describes a field of mass
$$
m^2 = \frac{a^2 e^2}{a-1},
\eqno(2.14)
$$
and $m^2 >0$ when $ a>1$. As long as $a>1$, this
system consists of a free massive degree of freedom
described by the field $\sigma$ such that harmonic
excitations propagate along the light cone described 
by the field $h$.

Moreover, the quantity $F = \epsilon^{\mu \nu} \partial_{\mu}
A_{\nu}$ obeys the same free massive Klein-Gordon
equation (2.13) satisifed by $\sigma$. This can be
shown by simplifying $F$ as
$$
\epsilon^{\mu \nu} \partial_{\mu} A_{\nu}
=- \frac{1}{ae} [ \epsilon^{\mu \nu}
\partial_{\mu} \partial_{\nu} \varphi
+ (1- a) \epsilon^{\mu \nu} \epsilon_{\nu}^{\tau}
\partial_{\mu} \partial_{\tau} \varphi
- a \epsilon^{\mu \nu} \epsilon_{\nu}^{\tau}
\partial_{\mu} \partial_{\tau} h ]
$$
$$
- \frac{1}{ae} [ ( 1-a) \Box \varphi -a \Box h ]
= \frac{ (a-1)}{ae} \Box \varphi
=- ae ( \varphi + h).
\eqno(2.15)
$$
When $\sigma = \varphi + h$ satisfies Klein-Gordon
equation (2.13), we find that
$$
( \Box + m^2) F =-ae ( \Box + m^2) (\varphi + h)
= -ae ( \Box + m^2) \sigma =0.
$$
This proves the claim.

\begin{center}
{\bf 3. Canonical Quantization of the Theory for $a>1$.}
\end{center}

The Lagrangian (2.5) can be put in the form
$$
{\cal L} =- \frac{1}{2} F_{01} F^{01} + \frac{1}{2}
( \partial_0 \varphi )^2 - \frac{1}{2} (\partial_1 \varphi)^2
+ e ( g^{0 \nu} - \epsilon^{0 \nu}) \partial_0 \varphi
A_{\nu} + e ( g^{1 \nu} - \epsilon^{1 \nu} )
\partial_1 \varphi A_{\nu} + \frac{1}{2} a e^2 A_{\mu} A^{\mu}.
\eqno(3.1)
$$
From the Lagrangian in this form, the canonical
momenta are found by calculating
$$
\pi_r (x,t) = \frac{\partial L}{\partial ( \partial_0 Q_r)}.
\eqno(3.2)
$$
Replacing $Q_r$ by $A_0$, $A_1$ and $\varphi$ respectively,
we obtain the momenta $\pi_0$, $\pi_1$ and $\pi$
$$
\pi_0 = \frac{\partial L}{\partial \dot{A}_0} =0,
\eqno(3.3)
$$
$$
\pi_1 = \frac{\partial L}{\partial \dot{A}_1}
= - ( \partial^0 A^1 - \partial^1 A^0)
=- F^{01} = F_{01},
\eqno(3.4)
$$
$$
\pi = \frac{\partial L}{\partial \dot{\varphi}}
= \partial_0 \varphi + e ( g_{0 \mu} - \epsilon_{0 \mu})
A^{\mu}.
\eqno(3.5)
$$
The Hamiltonian density and Hamiltonian can be
determined from the momenta and the Lagrangian density as
$$
{\cal H} = \pi \dot{\varphi} + \pi_1 \dot{A}_1 
+ \pi_0 \dot{A}_0 - {\cal L}
$$
$$
= \frac{1}{2} \pi_1^2 + \frac{1}{2} \pi^2 + \frac{1}{2}
( \partial_1 \varphi )^2 + \pi_1 \partial^1 A_0 - \frac{1}{2}
a e^2 A_{\mu} A^{\mu} + \frac{1}{2} e^2
( g_{0 \mu} - \epsilon_{0 \mu} )( g_{0 \nu} - \epsilon_{0 \nu})
A^{\mu} A^{\nu}
$$
$$
- e \pi ( g_{0 \mu} - \epsilon_{0 \mu}) A^{\mu} 
- e ( g_{1 \nu} - \epsilon_{1 \nu} ) \partial^1 \varphi A^{\nu}.
\eqno(3.6)
$$
From (3.3), we introduce the first class constraint
$\Omega_1 = \pi_0 \approx 0$ and incorporate $\Omega_1$ into 
the total Hamiltonian $H_T$ by means of a
Lagrange multiplier $\lambda_0 (x,t)$
$$
H_T = H + \int dx \, \lambda_0 \, \pi_0,
\eqno(3.7)
$$
where
$$
H = \int \, dx \, ( \frac{1}{2} \pi_1^2 + \frac{1}{2}
\pi_0^2 - A_0 \partial^1 \pi_1 + \frac{1}{2} 
( \partial_1 \varphi)^2 - \frac{1}{2} a e^2 A_{\mu} A^{\mu}
+ \frac{1}{2} e^2 ( g_{0 \mu} - \epsilon_{0 \mu} )
( g_{0 \nu} - \epsilon_{0 \nu}) A^{\mu} A^{\nu}
$$
$$
- e \pi ( g_{0 \mu} - \epsilon_{0 \mu}) A^{\mu}
- e ( g_{1 \nu} - \epsilon_{1 \nu}) \partial^1 \varphi A^{\nu}.
\eqno(3.8)
$$
Now it is required that the primary constraint
$\Omega_1$ be preserved in time under the action of the
Hamiltonian $H$,
$$
\dot{\pi}_0 = \{ \pi_0, H \},
\eqno(3.9)
$$
where these brackets denote the standard 
Poisson bracket defined as,
$$
\{ f_1 (y), f_2 (x) \} = \int \, d \tau \,
\sum_{j} [ \frac{\partial f_1 (y)}{\partial q_j (\tau)}
\frac{\partial f_2 (x)}{\partial p_j (\tau)}
- \frac{\partial f_1 (y)}{\partial p_j (\tau)}
\frac{\partial f_2 (x)}{\partial q_j (\tau)} ].
\eqno(3.10)
$$
This requirement leads to the existence of a
second-class constraint, namely,
$$
\Omega_2 \equiv \partial^1 \pi_1 + a e^2 A^0
- e^2 ( g_{0 \nu} - \epsilon_{0 \nu}) A^{\nu}
+ e \pi - e \partial^1 \varphi \approx 0.
\eqno(3.11)
$$
For the case here in which $a >1$, no new
constraints are generated by requiring the persistence 
in time of $\Omega_2$ in (3.11). Since the
Poisson bracket
$$
\{ \Omega_1 (y), \Omega_{2} (x) \} = \{
\pi_0, \partial^1 \pi_1 + a e^2 A_0 - e^2 A_0
- e^2 ( g_{01} - \epsilon_{01}) A^1 + e \pi
- e \partial^1 \varphi \}
$$
$$
= - (a^2 -1) e^2 \delta (y-x)
\eqno(3.12)
$$
does not vanish for $a>1$, the constraints are second-class.
Hence, requiring that $\dot{\Omega}_2 =0$ only acts to
determine the Lagrange multiplier $\lambda_0$ in (3.7).
The nonvanishing of the bracket implies that the local 
gauge invariance has been broken at the level
of the effective Lagrangian.

The matrix of Poisson brackets which is based on
the constraints $\Omega_{\alpha}$ is $2 \times 2$
and has the form
$$
\Delta_{\alpha \beta} (y,x) = \{ \Omega_{\alpha} (y),
\Omega_{\beta} (x) \} =
\left(
\begin{array}{cc}
0  &  - (a^2 -1) e^2 \delta (y-x)   \\
(a-1) e^2 \delta (y-x)  &   0       \\
\end{array}
\right).
\eqno(3.13)
$$
This is a nonsingular matrix, and its inverse
is required to evaluate the Dirac brackets for this case.
The inverse matrix is given by
$$
\Delta^{-1}_{\alpha \beta} (y,x) =
\left( 
\begin{array}{cc}
0   & \dss \frac{1}{e^2 (a-1)} \delta (y-x)   \\
- \dss \frac{1}{e^2 (a-1)} \delta (y-x)  &   0  \\
\end{array}  \right)
\eqno(3.14)
$$
It can be verified that $\Delta^{-1}_{\alpha \beta}$
satisfies the condition
$$
\int \, d \tau \, \Delta (y, \tau) \Delta^{-1} ( \tau, x)
= {\bf 1} \delta (y-x).
\eqno(3.15)
$$
The Dirac brackets can be evaluated by means of the
matrix elements of $\Delta^{-1}$ given that the canonically
conjugate pairs are $(\varphi, \pi)$, $(A_0, \pi_0)$ and
$(A_1 , \pi_1)$. Once these brackets are known, Dirac's
algorithm generates a quantization scheme. In terms
of the constraints $\Omega_{a}$, the Dirac bracket {\bf [5]}
is given by
$$
[ f_1 , f_2 ]_D = \{ f_1 , f_2 \} - \{ f_1 , \Omega_s \}
\Delta_{s s'}^{-1} \{ \Omega_{s'} , f_2 \}.
\eqno(3.16)
$$
For example, the following bracket yields
$$
[ A_1 (y), \pi_1 (x) ]_D = \int \, dz \, \delta (y-z)
\delta (z-x) - \int \, dz dz' ( \{ A_1 (y), \pi_0 (z) \}
\Delta_{12}^{-1} \{ \Omega_2 (z'), \pi_1 (x) \}
$$
$$
- \{ A_1 (y) , \Omega_2 (z) \} \Delta_{21}^{-1}
\{ \pi_0 (z') , \pi (x) \} )
= \delta (y-x),
$$
since both off-diagonal elements of $\Delta^{-1}$ are zero.

The canonical quantization of the theory is
achieved by abstracting the equal-time 
commutators from the corresponding Dirac brackets.
The quantum theory is obtained by taking the
commutation relations to correspond to
these new bracket relations. Thus the Dirac
brackets are replaced by commutators and a
multiplicative factor of $i$ is placed 
with what results on
the right-hand side. 
Nonvanishing equal time commutators are
presented here
$$
[ \varphi (y), \pi (x) ] = i \delta (y-x),
$$
$$
[ A_1 (y), \pi_1 (x) ] = i \delta (y-x),
$$
$$
[ A_0 (y), A_1 (x) ] = \frac{i}{e^2 (a-1)} 
\partial_y \delta (y-x),
$$
$$
[ A_0 (y), \varphi (x) ] = \frac{i}{e (a-1)}
\delta (y-x),
\eqno(3.17)
$$
$$
[ A_0 (y), \pi (x)] =- \frac{i}{a-1} 
\delta (y-x),
$$
$$
[ A_0 (y), \pi (x) ] = \frac{i}{e (a-1)}
\partial_y \delta (y-x).
$$

\begin{center}
{\bf 4. The Gauge-Noninvariant Theory For $a=1$}
\end{center}

Many of the commutators in (3.16) obtained for
$a>1$ become singular as $a$ approaches one
and the structure of the constraint $\Omega_2$ in 
(3.11) changes significantly when $a=1$.
The theory will be studied in more detail
for this case. The constraints become more
complicated and so to simplify we set $e=1$ and
also $a=1$ in the Lagrangian density.
It is given by
$$
{\cal L} = \frac{1}{2} ( \dot{\varphi}^2 -
\varphi^{'2} ) + (\dot{\varphi} + \varphi')
(A_0 - A_1) + \frac{1}{2} ( \dot{A}_1 - A_{0}')^2
+ \frac{1}{2} ( A_0^2 - A_1^2).
\eqno(4.1)
$$
To simplify ${\cal L}$, it has been expanded out in detail
and dot and prime denote time and space derivatives,
respectively. The terms in (4.1) have interpretations.
The first term corresponds to a massless boson
the second represents the chiral coupling of
$\varphi$ to the electromagnetic field $A_{\mu}$,
the third term is the kinetic energy of the 
electromagnetic field, and the last term is
associated with the mass for the vector particle.

The canonical momenta are determined to be
$$
\pi_0 = \frac{\partial L}{\partial \dot{A}_0}=0,
$$
$$
\pi_1 = \frac{\partial L}{\partial \dot{A}_1} =
\dot{A}_1 - A_{0}',
$$
$$
\pi = \frac{\partial L}{\partial \dot{\varphi}}
= \dot{\varphi} + A_0 - A_1.
$$
The Hamiltonian density can be determined using
these momenta from
$$
{\cal H} = \pi \dot{\varphi} + \pi_1 \dot{A}_1
+ \pi_0 \dot{A}_0 - {\cal L}.
$$
It is determined to be
$$
{\cal H} = \frac{1}{2} \pi^2 + \frac{1}{2} \pi_1^2 +
\frac{1}{2} \varphi^{'2} + \pi_1 A_0 ' +
( \pi + \varphi' + A_1)(A_1 - A_0),
\eqno(4.2)
$$
and the Hamiltonian $H$ is the integral of
${\cal H}$ over the space variable. 
The canonically conjugate pairs can
then be summarized as $(\varphi, \pi)$, $(A_0, \pi_0)$
and $(A_1, \pi_1)$. The Lagrangian density in
(4.1) possesses the following four second-class
constraints
$$
\begin{array}{c}
\Omega_1 = \pi_0 \approx 0,    \\
   \\
\Omega_2 = \pi_1' + \varphi ' + \pi +A_1 \approx 0, \\
   \\
\Omega_3 = \pi_1 \approx 0,    \\
   \\
\Omega_4 = - \pi - \varphi' -2 A_1 + A_0 \approx 0.
\end{array}
\eqno(4.3)
$$
In (4.3), $\Omega_1$ is a primary constraint
and $\Omega_2$, $\Omega_3$ and $\Omega_4$ are
secondary constraints.

To prove this we proceed as follows.
The momentum $\pi_0$ is seen to vanish hence
$\pi_0 \approx 0$ is a primary constraint.
Now it is required that this constraint be
invariant under the action of the Hamiltonian.
By calculating the Poisson bracket
$\{ \pi_0 (y), H (x) \}$ and requiring that this
be zero generates a new constraint, $\Omega_2$.
Proceeding in a similar way, the remaining two
constraints are obtained. Using the constraint 
$\Omega_{\alpha}$, the matrix of Poisson brackets can be
calculated using (3.10) explicitly, and it is given by
$$
\Delta_{\alpha \beta} (y,x) =
\left(
\begin{array}{cccc}
0 &  0  &  0  & - \delta (y-x)   \\
0 &  0  &  \delta (y-x) &  0     \\
0 & - \delta (y-x)  & 0  &  2 \delta(y-x)    \\
\delta (y-x) &  0  & - 2 \delta (y-x)  & 2 \partial_y \delta (y-x) \\
\end{array}  \right)
\eqno(4.4)
$$
This matrix is nonsingular and has an inverse
$\Delta_{\alpha \beta}^{-1}$ which satisfies (3.15)
and can be
used to calculate the Dirac bracket by means
of (3.15). Once these are obtained, Dirac's
algorithm for quantization discussed before 
can be applied.

Following the same procedure, the non-vanishing 
equal-time commutators obtained by the
quantization of this system are presented below

$$
[ A_0 (y), \varphi (x) ] = [ A_1 (y), \varphi (x)]
= [ \varphi (y), \pi (x)] = i \delta (y-x),
$$
$$
[A_0 (y), \pi (x) ] = [ A_1 (y), \pi (x) ] =-i \partial_y
\delta (y-x),
\eqno(4.5)
$$
$$
[ A_0 (y), A_0 (x) ] = [ A_0 (y), A_1 (x) ]
= [ A_1 (y), A_1 (x) ]
= 2 i \partial_y \delta (y-x).
$$

\begin{center}
{\bf 5. The Gauge Invariant Theory.}
\end{center}

In constructing a gauge-invariant model
corresponding to the Lagrangian in (4.1),
a type of Wess-Zumino term is calculated {\bf [10]}.
To do this, the actual Hilbert space of the
theory is expanded to include a new field, 
which we call $\vartheta$.
This is done by redefining the fields $\varphi$ and
$A^{\mu}$ in the original Lagrangian density as {\bf [18]}
$$
\varphi \rightarrow \varphi - \vartheta,
\qquad
A^{\mu} \rightarrow A^{\mu} + \partial^{\mu} \, \vartheta.
\eqno(5.1)
$$
Under this replacement, ${\cal L}$ is mapped
into ${\cal L}_T$ given by
$$
{\cal L}_T = \frac{1}{2} ( \dot{\varphi}^2 -
\varphi^{'2} ) + ( \dot{\varphi} + \varphi')
( A_0 - A_1) + \frac{1}{2} ( \dot{A}_1 - A_0')^2
+ \frac{1}{2} (A_0^2 - A_1^2)
$$
$$
+ \varphi' \dot{\vartheta} - \dot{\varphi} \vartheta' +
\dot{\vartheta} A_1 - \vartheta ' A_0
\eqno(5.2)
$$
$$
= {\cal L} + {\cal L}_{\vartheta}.
$$
Here, we have defined
$$
{\cal L}_{\vartheta} = \varphi ' \dot{\vartheta}
- \dot{\varphi} \vartheta' + \dot{\vartheta} A_1 -
\vartheta' A_0.
\eqno(5.3)
$$
Since the total action is an integral of (5.2) over $(x,t)$,
the first two terms in (5.3) could be eliminated
from the action using integration by parts.
However, they have an effect on the structure of 
the constraints and should be retained here. The 
constraint structure is very important as far as 
quantization is concerned, in particular, as far
as path integral quantization is concerned when
gauge constraints must be invoked. In fact,
${\cal L}_T$ describes a gauge-invariant theory.

The Euler-Lagrange equations obtained from
${\cal L}_T$ including the terms $\varphi' \dot{\vartheta}
- \dot{\varphi} \vartheta'$ in ${\cal L}_{\vartheta}$ are 
identical to the Lagrange equations without 
these terms and are given by
$$
\begin{array}{c}
\ddot{\varphi} - \varphi'' = \dot{A}_1 - A_0 ' - \dot{A}_0
+ A_1 ',   \\
   \\
\ddot A_1 - \dot{A}_0 ' = \dot{\vartheta} - \dot{\varphi}
- \varphi' - A_1,   \\
   \\
\dot{A}_1 ' - A_0 '' = \vartheta ' - A_0 - \dot{\varphi} -
\varphi ',    \\
   \\
\dot{A}_1 - A_0 ' =0.
\end{array}
\eqno(5.4)
$$
Using ${\cal L}_T$ in (5.2), the canonical
momenta for the gauge-invariant theory are calculated 
to be
$$
\begin{array}{c}
\pi_0 = \dss \frac{\partial L_T}{\partial \dot{A}_0} =0,  \\
  \\
\pi_{\vartheta} = \dss \frac{\partial L_T}
{\partial \dot{\vartheta}} = A_1 + \varphi',   \\
  \\
\pi_1 = \dss \frac{\partial L_T}{\partial \dot{A}_1}
= \dot{A}_1 - A_0 ',  \\
  \\
\pi = \dss \frac{\partial L_T}{\partial \dot{\varphi}}
= \dot{\varphi} + A_0 - A_1 - \vartheta '.
\end{array}
\eqno(5.5)
$$
Thus the theory possesses two primary constraints,
each independent of velocity terms
$$
\psi_1 = \pi_0 \approx 0, \qquad
\psi_2 = \pi_{\vartheta} -A_1 - \varphi ' \approx 0. 
\eqno(5.6)
$$
Only the momenta $\pi$ and $\pi_1$ in (5.5) involve time derivatives,
and the time derivatives can be solved for explicitly
$$
\dot{A}_1 = \pi_1 + A_0 ',
\qquad
\dot{\varphi} = \pi - A_0 + A_1 + \vartheta '.
\eqno(5.7)
$$
The canonical Hamiltonian density can be
calculated from these as
$$
{\cal H}_T = \pi \dot{\varphi} + \pi_{\vartheta} \dot{\vartheta}
+ \pi_1 \dot{A}_1 + \pi_0 \dot{A}_0 - {\cal L}_T.
\eqno(5.8)
$$
Using (5.5) and (5.7), ${\cal H}_T$ is found to be
$$
{\cal H}_T = \frac{1}{2} (\pi^2 + \pi_1^2)
+ \frac{1}{2} ( \varphi^{'2} + \vartheta^{'2})
+ \pi_1 A_0' + (\pi + \varphi' + A_1 + \vartheta')(A_1 - A_0)
+ \vartheta' A_0 + \pi \vartheta'.
\eqno(5.9)
$$
All of the velocities have been eliminated in 
obtaining (5.9), and only derivatives with respect to 
the spatial coordinates remain, even as far as
the $\vartheta$ field is concerned.

Again, the primary constraints can be included in
the canonical Hamiltonian density by making use
of Lagrange multipliers $\lambda_0$ and $\lambda_1$
as follows
$$
{\cal H}_E = {\cal H}_T + \lambda_0 \pi_0
+ \lambda_1 ( \pi_0 - A_1 - \varphi ')
$$
$$
= \frac{1}{2} ( \pi^2 + \pi_1^2 ) +
\frac{1}{2} ( \varphi^{'2} + \vartheta^{'2} ) +
\pi_1 A_0' + (\pi + \varphi' + A_1 + \vartheta')
(A_1 - A_0) + \vartheta' A_0 + \pi \vartheta'
\eqno(5.10)
$$
$$
+ \pi_0 \lambda_0 + (\pi_{\vartheta} -A_1 - \varphi ' ) \lambda_1.
$$
The total Hamiltonian is given by the integral of ${\cal H}_E$
over $x$. From the total Hamiltonian the set of Hamilton's
equations can be obtained. This will be done since they can be used 
as an alternative way to determine the evolution of 
the constraints under the action of the Hamiltonian.
By differentiating the Hamiltonian, we have
$$
\begin{array}{cc}
\dot{\varphi} = \dss \frac{\partial H_E}{\partial \pi} = \pi + A_1 - A_0 + \vartheta',
&
- \dot{\pi} = \dss \frac{\partial H_E}{\partial \varphi}
=- \varphi'' - A_1' +A_0' + \lambda_1',     \\
  \\
\dot{A}_0 = \dss \frac{\partial H_E}{\partial \pi_0} = \lambda_0,
&
- \dot{\pi}_0 = \dss \frac{\partial H_E}{\partial A_0} 
=- \pi_1' - \pi - \varphi' -A_1,             \\
  \\
\dot{A}_1 = \dss \frac{\partial H_E}{\partial \pi_1} = \pi_1 + A_0',
&
- \dot{\pi}_1 = \dss \frac{\partial H_E}{\partial A_1}
= \pi + \varphi' +2 A_1 + \vartheta' - A_0 - \lambda_1,  \\
   \\
\dot{\vartheta} = \dss \frac{\partial H_E}{\partial \pi_{\vartheta}} 
= \lambda_1, 
&
- \dot{\pi}_{\vartheta} = \dss \frac{\partial H_E}{\partial \vartheta}
= - \vartheta'' - A_1 ' - \pi',    \\
  \\
\dot{\lambda}_0 = \dss \frac{\partial H_E}{\partial p_{\lambda_0}} =0,
&
- \dot{p}_{\lambda_0} = \dss \frac{\partial H_E}{\partial \lambda_0} 
= \pi_0,      \\
   \\
\dot{\lambda}_1 = \dss \frac{\partial H_E}{\partial p_{\lambda_1}} =0,
&
- \dot{p}_{\lambda_1} = \dss \frac{\partial H_E}{\partial \lambda_1}
= \pi_{\vartheta} - A_1 - \varphi'.   \\
\end{array}
\eqno(5.11)
$$

The two primary constraints are $\psi_1$ and $\psi_2$,
and it is required that these constraints be preserved in time.
Demanding that the primary constraint $\psi_1$ be preserved
in time, a secondary constraint is obtained. 
Using Hamilton's equations, since $\dot{\pi}_0$ is now known,
$$
\{ \psi_1 , H_E \} = \dot{\pi}_0 = \pi_1' + \pi + \varphi' + A_1 \approx 0.
\eqno(5.12)
$$
Thus, (5.12) gives a third constraint
$$
\psi_3 = \pi_1' +\pi + \varphi' + A_1 \approx 0.
\eqno(5.13)
$$
The constraint $\psi_3$ leads in turn to a fourth
constraint $\psi_4$. Using Hamilton's equations (5.14),
this is found by evaluating
$$
\{ \dot{\pi}_0, H_E \} = \ddot{\pi}_0 
= \dot{\pi}_1' + \dot{\pi} + \dot{\varphi}' + \dot{A}_1
= \pi_1 + \lambda_1 '.
\eqno(5.14)
$$
The preservation of $\psi_2$ and $\psi_4$ in time
do not yield further constraints if we make $\lambda_1$ 
independent of $x$. Thus, the theory is seen to
possess four constraints which are summarized below
$$
\begin{array}{c}
\psi_1 = \pi_0 \approx 0,  \\
  \\
\psi_2 = \pi_{\vartheta} - A_1 - \varphi' \approx 0,  \\
  \\
\psi_3 = \pi_1' + \pi + \varphi' + A_1 \approx 0,     \\
  \\
\psi_4 = \pi_1 \approx 0.  \\
\end{array}
\eqno(5.15)
$$
The conjugate pairs for the system now read
$(\varphi, \pi)$, $(A_0, \pi_0)$, $(A_1, \pi_1)$,
$( \vartheta, \pi_{\vartheta})$, $(\lambda_0,
p_{\lambda_0})$ and $(\lambda_1, p_{\lambda_1})$.
A current can be defined in this case which is
conserved and it is given by $J^{\nu}
= \partial_{\mu} F^{\mu \nu}$. Using (2.7) with $a=1$,
and equations (5.4), we find that
$$
-\partial_{\nu} ( \partial_{\mu} F^{\mu \nu} )
= ( g^{\nu \alpha} - \epsilon^{\nu \alpha})
\partial_{\nu} \partial_{\alpha} \varphi +
\partial_{\nu} A^{\nu}
= \dot{A}_0 - A_1 ' + \ddot{\varphi} - \varphi''
$$
$$
= \dot{A}_0 - A_1' + \dot{A}_1 - A_0' - \dot{A}_0 + A_1'
= \dot{A}_1 - A_0' =0.
$$
Thus $\partial_{\nu} J^{\nu} =0$
 and so the gauge-invariant theory is
nonanomalous.

The next step is to work out the matrix of
Poisson brackets, which is $4 \times 4$ in this case
for the constraints $\psi_a$. It is found to
be a singular matrix with a row and a column
of zeros appearing in the matrix. This implies 
that the set of constraints $\psi_a$ form a 
set of first class constraints, and the theory
described by the Lagrangian is a gauge-invariant
theory.

If the theory is going to be quantized using Dirac's
procedure, the first-class constraints of the
theory must be converted into second class constraints.
To do this, some additional constraints are to be
imposed arbitrarily on the system. These are 
what is referred to as a set of gauge-fixing 
conditions. Suppose we require that the $\vartheta$ field 
satisfy the special condition given by requiring
$$
\partial^{\mu} \vartheta =0.
\eqno(5.16)
$$
This condition can be satisfied by taking the
following pair of equations to hold
simultaneously as the two gauge conditions
$$
\dot{\vartheta} =0,
\qquad
- \vartheta ' =0.
\eqno(5.17)
$$
Thus, using the fact that $\dot{\pi}_1 \approx 0$,
the remaining constraint is taken to be
$$
\dot{\vartheta} =- \pi - \varphi' - 2 A_1 + A_0 - \vartheta'
\approx 0,
$$
and the total set of six constraints for the theory in this
form is summarized here
$$
\chi_1 = \psi_1 = \pi_0 \approx 0,
$$
$$
\chi_2 = \psi_2 = \pi_{\vartheta} -A_1 - \varphi ' \approx 0,
$$
$$
\chi_3 = \psi_3 = \pi_1 ' + \pi + \varphi' +A_1 \approx 0,
$$
$$
\chi_4 = \psi_4 = \pi_1 \approx 0,
\eqno(5.18)
$$
$$
\chi_5 = G_1 = - \vartheta ' \approx 0,
$$
$$
\chi_6 = G_2 =- \pi - \varphi' -2 A_1 + A_0 + \vartheta' \approx 0.
$$
Due to the presence of the new constraint $\xi_6$,
there exists a coupling between $\xi_1$ and $\xi_6$
in the Poisson brackets,
since the variables $A_0$ and $\pi_0$ occur in a
canonical pair in the matrix of Poisson brackets.

From this collection of Poisson brackets, the
corresponding matrix of Poisson brackets based
on the six constraints $\chi_{a}$ can be
written down in the following form
$$
\Delta_{\alpha \beta} (y,x)  
$$
$$
= \left(
\begin{array}{cccccc}
0 & 0 & 0 & 0 & 0  & - \delta( y-x)  \\
0 & 0 & 0 & - \delta (y-x) & \partial_y \delta (y-x) &  0  \\
0 & 0 & 0 & \delta (y-x)   &  0  &  0  \\
0 & 0 & 0 & \delta (y-x)   &  0  &  0  \\
0 & \delta (y-x) & - \delta (y-x) & 0 &  0  & 2 \delta (y-x)  \\
0  & - \partial_y \delta (y-x) &  0  &  0  & 0  &  0  \\
\delta (y-x) &  0  &  0  &  -2 \delta (y-x) & 0  & 2 \partial_y \delta (y-x)  \\
\end{array}
\right)
\eqno(5.19)
$$
This matrix is nonsingular and an inverse $\Delta_{\alpha \beta}^{-1}$
can be calculated which satisfies (3.15).
Using the inverse matrix, the Dirac brackets
can be calculated and the theory can be
quantized in the same way as before using (3.16).
This process reproduces the commutators
already given in (4.5) and in addition
generates a few additional commutators
which pertain to the new $\vartheta$ field, namely
$$
[ \vartheta (y), \pi_{\vartheta} (x) ] = 2 i \delta (y-x),
\qquad
[ \pi_{\vartheta} (y), \varphi (x) ] =-i \delta (y-x),
$$
$$
[ A_0 (y), \pi_{\vartheta} (x) ]
= 2 [ \pi_{\vartheta} (y), \pi (x)] 
= - [ \pi_{\vartheta} (y) , \pi_{\vartheta} (x) ]
= 2 i \partial_y \delta (y-x).
\eqno(5.20)
$$

\begin{center}
{\bf 6. Summary and Further Ideas.}
\end{center}

The constraint structure for this model has
been examined. The method of Dirac brackets
provides a well defined strategy for finding
a canonical quantization. This has
been done for two regimes of the arbitrary
renormalization parameter in the model.
There are other ways to quantize classical
systems, and for comparison and future work,
we consider the path integral approach.
The path integral also
provides a means of quantizing a theory.
Since the introduction of the $\vartheta$ term
has led to a gauge invariant theory, it
would seem appropriate to apply that here.
To do this, the two constraints $\chi_1$
and $\chi_2$ in (5.18) are taken with the gauge
fixing conditions $G_1$ and $G_2$ and the
Poisson brackets $\{ \chi_a , G_c \}$ are
evaluated then put in a $2 \times 2$ matrix.
The determinant can be evaluated, and so
the transition amplitude can be
expressed in the form of a path integral
$$
{\cal A} = \int \prod_t \prod_{c=1}^2 \delta (\chi_c)
\det | \{ \chi_a, G_b \}| \frac{D  \pi D \varphi}{(2 \pi)^2}
\frac{D \pi_0 D A_0}{(2 \pi)^2 } \frac{D \pi_1 D A_1}{(2 \pi)^2}
\frac{D \pi_{\vartheta} D \vartheta}{(2 \pi)^2}
$$
$$
\cdot \frac{D \lambda_c (t)}{2 \pi} \exp
\{ i \int_{t'}^{t''} [ \pi \dot{\varphi} + \pi_0 \dot{A}_0
+ \pi_1 \dot{A}_1 + \pi_{\vartheta} \dot{\vartheta}  - H_T
- \sum_{a=1}^{2} \lambda_{a} \chi_a ] \}.
$$
The gauge-invariant version of the model
can be written as a system that possesses BRST
symmetry. This symmetry can be thought
of as a generalized gauge invariance.
Quantization can also be done in this way as well.

\begin{center}
{\bf References.}
\end{center}

\noindent
$[1]$ J. Schwinger, Phys. Rev. {\bf 128}, 2435, (1962).  \\
$[2]$ R. Jackiw and R. Rajamaran, Phys. Rev. Lett. {\bf 54},
1219, (1985).     \\
$[3]$ R. Rajaraman, Phys. Letts {\bf B 154}, 305, (1985) \\
$[4]$ U. Kulshreshtha, D. Kulshreshtha and H. M\"{u}ller-Kirsten,
Can. J. Phys. {\bf 72}, 639, (1994).  \\
$[5]$ P. A. M. Dirac, Lectures on Quantum Mechanics, Dover
Publications, NY, (2001).  \\
$[6]$ M. Henneaux and C. Teitelboim, Quantization of Gauge
Systems, Princeton University Press, Princeton, NJ, (1992).  \\
$[7]$ L. D. Faddeev and A. A. Slavnov, Gauge Fields, Introduction 
to Quantum Theory, Benjamin-Cummings Publishing Company, 1980.  \\ 
$[8]$ A. A. Deriglazov and Z. Kuznetsova, Phys. Letts. {\bf B 646},
47, (2007).  \\
$[9]$ I. Cortese and J. A. Garcia, Phys. Letts. {\bf A 358},
327, (2006).   \\
$[10]$ J. Wess and B. Zumino, Phys. Lett. {\bf B 37}, 95, (1971).  \\
$[11]$ C. Bechi, A. Rouet and R. Stora, Phys. Lett. {\bf B 52},
344, (1974).   \\
$[12]$ O. Babelon, F. A. Schaposnik and C. M. Viallet,
Phys. Letts. {\bf B 177}, 385, (1986).   \\
$[13]$ K. Harada and I. Tsutsui, Phys. Letts. {B 113},
385, (1987).    \\
$[14]$ M. S. Marinov, Phys. Reports {\bf 60}, 1, (1980).  \\
$[15]$ R. A. Bertlmann, Anomalies in Quantum Field Theory,
Clarendon Press, Oxford, (1996).  \\
$[16]$ L. H. Ryder, Quantum Field Theory, Cambridge University
Press (1988).  \\
$[17]$ H. Girotti, H. J. Rothe and K. D. Rothe, Phys.
Rev. {\bf D 33}, 514, (1986).  \\
$[18]$ E. Stueckelberg, Helv. Phys. Acta. {\bf 15}, 52, (1941),
Helv. Phys. Acta. {\bf 30}, 209, (1957).  \\
\end{document}